\documentclass[preprint]{aastex631}

\received{February 2, 2022}
\revised{March 15, 2022}
\accepted{April 1, 2022}

\shorttitle{BA$_{14}$ in Close-Up}
\shortauthors{Kareta et al.}
\graphicspath{{./}{figures/}}
\newcommand{\edits}[1]{\textcolor{black}{#1}}

\begin{document}

\title{Near-Infrared Spectroscopy Of The Nucleus Of Low-Activity \\
Comet P/2016 BA$_{14}$ During Its 2016 Close Approach}




\correspondingauthor{Theodore Kareta}
\email{tkareta@lowell.edu}

\author[0000-0003-1008-7499]{Theodore Kareta}
\affil{Lowell Observatory \\
Flagstaff, AZ, USA}
\affil{Lunar and Planetary Laboratory \\
University of Arizona \\
Tucson, AZ, USA}

\author{Vishnu Reddy}
\affil{Lunar and Planetary Laboratory \\
University of Arizona \\
Tucson, AZ, USA}

\author{Juan A. Sanchez}
\affil{Planetary Science Institute \\
Tucson, AZ, USA}

\author{Walter M. Harris}
\affil{Lunar and Planetary Laboratory \\
University of Arizona \\
Tucson, AZ, USA}



\begin{abstract}

The Near-Earth Comet P/2016 BA$_{14}$ (PanSTARRS) is a slow-rotatating nearly-dormant object, a likely dynamical twin of 252P/LINEAR, and was recently shown to have a mid-infrared spectrum very dissimilar to other comets. BA$_{14}$ also recently selected one of the back-up targets for the ESA's \textit{Comet Interceptor}, so a clearer understanding of BA$_{14}$'s modern properties would not just improve our understanding of how comets go dormant, but could also aid planning for a potential spacecraft visit. We present observations of BA$_{14}$ taken on two dates during its 2016 Earth close approach with the NASA Infrared Telescope Facility, both of which are consistent with direct observations of its nucleus. The reflectance spectrum of BA$_{14}$ is similar to 67P/Churyumov-Gerasimenko, albeit highly phase-reddened. Thermal emission contaminates the reflectance spectrum at longer wavelengths, which we correct with a new Markov Chain Monte Carlo thermal modeling code. The models suggest $BA_{14}$'s visible geometric albedo is $p_V=0.01-0.03$, consistent with radar observations, its beaming parameter is typical for NEOs observed in its geometry, and its reflectrance spectrum is red and linear throughout H and K band. It appears very much like a ``normal" comet nucleus, despite its mid-infrared oddities. A slow loss of fine grains as the object's activity diminished might help to reconcile some of the lines of evidence, and we discuss other possibilities. A spacecraft flyby past BA$_{14}$ could get closer to the nucleus than with a more active target, and we highlight some science questions that could be addressed with a visit to a (nearly-)dormant comet.

\end{abstract}



\section{Introduction} \label{sec:intro}
The study of low activity or fully dormant comets is critical to understanding how comets age and deplete their volatiles as well as learning how to identify them among other low-albedo small bodies in the inner Solar System. The problem is identifying them. While some orbits might allow for more definitive conclusions about whether or not a particular object is a dormant comet (e.g. \edits{lower} Tisserand parameters), many objects cannot be so conclusively identified as such without additional information. One scenario where a cometary origin can be more clearly identified is if a candidate dormant comet is dynamically linked to a traditionally active comet, such as with the dormant comet 2003 EH$_1$ and active comet 96P/Machholz \citep{2004AJ....127.3018J, 2005Icar..179..139W}. (EH$_1$ is also the parent of the Quadrantids meteor shower, so their dynamical relationship wasn't the only clue.) Dynamically similar or linked objects have likely experienced similar thermal conditions over the past few thousand years due to their similar orbits, and if they are genetically related (e.g. descended from a common progenitor object), one can assume that their bulk compositions are quite similar as well. The problem is then to constrain which scenarios can explain their modern differences so as to better understand how comets age and turn off more directly than would otherwise be possible. When did they split, and why are they so different now? Studies of dormant-and-not-dormant comet pairs or groups are thus of great interest to those studying the life cycles of comets in particular and the divide between comets and asteroids in general.

The close approach of comet 252P/LINEAR (hereafter just '252P') to the Earth on March 21st, 2016 was followed closely $\sim$ 26 hours later by the approach of P/2016 BA$_{14}$ (PANSTARRS, hereafter just 'BA$_{14}$') on the 22nd. BA$_{14}$ had only been discovered in late January of that year \citep{2016MPEC....B...79L} as an apparently inactive Near-Earth Object (NEO) on the orbit of a Jupiter Family Comet extremely similar to that of 252P. Observations of BA$_{14}$ prior to the close approach revealed that it too was undergoing cometary mass-loss (first reported by \citealt{CBET_4257}), and it was subsequently given the cometary P/ designation completing its modern name \citep{2016MPEC....F...59N}. Post-encounter, 252P has a perihelion distance of $q=0.996$ AU and an eccentricity of $e=0.673$, while BA$_{14}$ has $q=1.008$ AU and $e=0.666$.

The two comets, despite their similar orbits, appear to be quite different. \citet{2017AJ....154..136L} characterized 252P with the Hubble Space Telescope and both 252P and BA$_{14}$ with the Lowell Discovery Telescope (LDT, then named the Discovery Channel Telescope) during and around their close approach, and found that while 252P is small ($D\sim0.6$km) and was very active for its size (a water-derived active fraction of 40 to more than 100\%), BA$_{14}$'s active fraction was on the order of $\sim0.01\%$ or so, or a factor of $\sim10^4$ less, despite its larger nucleus. \citet{2017AJ....154..136L} estimated BA$_{14}$'s nucleus to be approximately $D\sim1$km in size based on its JPL Horizon's reported absolute magnitude of $H_V=19.2$ and an assumed albedo of $4\%$. Radar observations presented by \citet{2016DPS....4821905N} found the size of the object to likely be more than a kilometer in diameter (implying an optical albedo of $\sim 0.03$ or less) and an extremely slow rotation period of $\sim40$ hours with hints of an angular and blocky surface. While \citet{2019MNRAS.484.1347H} did not detect gas emission from BA$_{14}$, \citet{2017AJ....154..136L} did detect CN emission on April 17th, 2016 consistent with a production rate of $Q(CN) = (1.4\pm0.1) \times 10^{22} mol/s$. The visible reflectivity of BA$_{14}$'s nucleus or coma is minimally constrained, with \citet{2017AJ....154..136L} only observing BA$_{14}$ in a single non-gas filter (r') and \citet{2019MNRAS.484.1347H} obtaining a red slope beyond $\sim0.58\mu{m}$. Their results hinted at a possible \textit{blue} slope at shorter wavelengths ($<0.54\mu{m}$) but calibration issues hampered their ability to ascertain how reliable that measurement was.

Perhaps most relevant to the current study is the recent work of \citet{2021Icar..36314425O} which presented and analyzed mid-infrared observations of BA$_{14}$ just prior to its close approach. Those authors argued that their observations were dominated by emission from BA$_{14}$'s nucleus as opposed to from grains in its coma. The direct detection of cometary nuclei during close approaches is possible due to the larger apparent area of the sky that the coma is spread over, which in this case would be assisted by the overall minimal activity of the object. A similar scenario of somewhat-unplanned direct mid-infrared observations of the nucleus of a comet during close approach is described in the seminal \citealt{1985Icar...62...97H}. \citet{2021Icar..36314425O} present mid-infrared photometry and low-resolution spectra of BA$_{14}$ that do not look similar to typical mid-infrared observatons of cometary comae (see, e.g., \citealt{2009P&SS...57.1133K}) or the nuclei of more active comets observed at larger heliocentric distances (see, e.g., \citealt{2017Icar..284..344K}). Instead of an emission spectrum dominated by emission features from silicates (perhaps most notably the 10-$\mu{m}$ excess), the retrieved spectrum appears dominated instead by absorption features due to phyllosilicates and organics. While \citet{Lisse} did find evidence for phyllosilicates in the ejecta of the Deep Impact experiment \citep{2005Sci...310..258A}, the dominant materials (by areal fraction in the retrieved spectral modeling, at least) were that of amorphous and crystalline silicates. \citet{2021Icar..36314425O} argue that if they did indeed observe the nucleus of BA$_{14}$, then a phyllosilicate-rich surface might be common surface composition for comets, or at least those on the verge of going dormant. Considering the lack of information about the surface of BA$_{14}$ at the wavelengths where comet nuclei are increasingly able to be studied (visible and near-infared), it is challenging to ascertain which of these scenarios is more likely, or if BA$_{14}$ is an outlier for a reason yet to be considered.

We thus have \edits{one} primary question we aim to address in this study \edits{and a second that will require additional work at a latter date}. Primarily, what is the actual nature of the surface of BA$_{14}$ and how could it inform our understanding of the surfaces of low-activity comets and how comets go dormant in general? \edits{Secondarily, what is the relationship between BA$_{14}$ and 252P, and what would explain their differences if they are genetically related? We do not present new observations of 252P or address this question in great detail in this work, but highlight part of the path forward at the end of the Section \ref{sec:summary}.} It is particularly critical to address these questions now, as BA$_{14}$ has recently been selected as a potential backup target for the ESA's \textit{Comet Interceptor} mission \citep{2019NatCo..10.5418S, 2020RNAAS...4...21S}. It is critical to understand these two objects in as much detail now such that the best-informed choice can be made nearer to the end of the decade when the Comet Interceptor team is choosing a target.

In this work, we present low-resolution near-infared ($0.7-2.5\mu{m}$) spectra of BA$_{14}$ taken with SpeX \citep{2003PASP..115..362R} on the NASA Infrared Telescope Facility (IRTF) from approximately 11 and 14 days prior to the observations of \citet{2021Icar..36314425O} to attempt to diagnose the surface reflectivity and albedo of this object and place it in context among the broader population of comets whose nuclei have been studied directly. In Section \ref{sec:obs}, we present a journal of observations and describe our data reduction procedure. In Section \ref{sec:results}, we present and describe the reflectance spectra we obtained of BA$_{14}$ on March 7th and 10th UTC and make comparisons to other comets observed at the same wavelengths. We also present spatial profiles of BA$_{14}$ at visible and near-infrared wavelengths for context about any possible coma contamination. In Section \ref{sec:thermal}, we describe and utilize our efforts to remove the thermal excess from the longer wavelength portions of our spectrum to analyze the underlying reflectance there and to constrain its albedo. Finally, in Section \ref{sec:discussion}, we synthesize all of these results and those in the literature to better understand how to interpret the properties of P/2016 BA$_{14}$ (PanSTARRS) and address what potential spacecraft exploration could do for the target.

\section{Observations} \label{sec:obs}

P/2016 BA$_{14}$ (PanSTARRS) was observed on March 7 and March 10 2016 UTC with the SpeX instrument \citep{2003PASP..115..362R} on the NASA Infrared Telescope Facility (IRTF) as part of an observational program centered on time-sensitive observations of small and/or close-approaching NEOs. The observational details are summarized in Table \ref{tab:observations}. All observations were obtained with the low-resolution 'Prism' mode on SpeX, providing simultaneous coverage from $0.7-2.5\mu{m}$ at an effective resolution of $R\sim100$ with a 0.8"-wide slit. Observations of the target were 'bookended' by observations of a G-type star nearby on the sky at similar airmass for proper telluric correction, and observations of a proper solar analog star (SAO 93936 on both nights) was later observed near 1.0 airmass for further slope correction of the final output spectrum. \edits{Observations of the target and calibration stars were all obtained at the local parallactic angle to mitigate wavelength-dependent slit losses.} The observational circumstances were such that all observations of BA$_{14}$ were between airmasses of $1.55-1.70$. The reduction, extraction, and telluric correction of the spectra were accomplished using the IDL-based 'spextool' package \citep{2004PASP..116..362C} using the default (optimal) settings for an unresolved point-source, while the final solar analog correction was accomplished using a custom-written script in Python.

\subsection{Comparison of Visible and Near-IR PSFs}
\begin{figure}[ht!]
\plotone{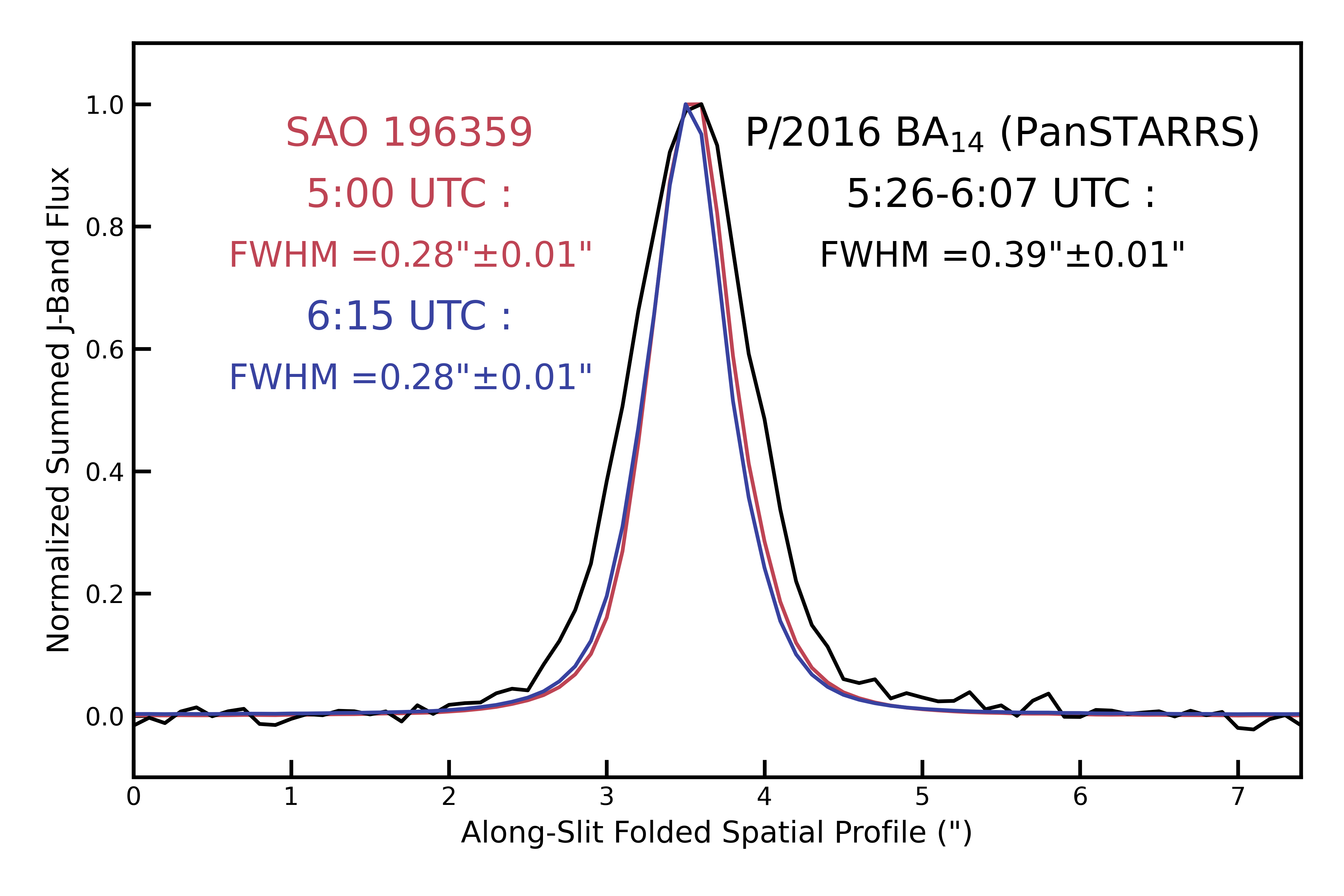}
\caption{The along-slit spatial profiles are shown for P/2016 BA$_{14}$ (PanSTARRS) as well as the telluric star observed immediately before and afterwards as obtained on 2016 March 7. The shown profiles are medians over the wavelengths corresponding to J-band ($1.17-1.33\mu{m}$). The visible seeing was quite good ($\sim0.51\arcsec$) and the resulting profiles are quite narrow, but the profile of P/2016 BA$_{14}$ is still significantly wider than the star was before or after. Worse seeing (such as that obtained on 2016 March 10) would not have shown this clear hint for activity, however weak it might be. See text for more details.}
\label{fig:psfs}
\end{figure}

While the reduction was completed shortly after observations, the SpeX observations are (thankfully) available on the IRTF Legacy Archive\footnote{http://irtfdata.ifa.hawaii.edu/search/}, and we present a comparison of the PSFs of local telluric star observed before and after observations of the target were taken to assess whether or not any dust coma might be visible in Figure \ref{fig:psfs}. The data shown are from 2016 March 7, where seeing was much better ($\sim0.51\arcsec$ at visible wavelengths \edits{as measured by the visible wavelength guide camera}). The shown spatial profiles were derived by combining the "A" and "B" profiles after doing an A-B subtraction on paired sequetial frames (labeled as a "folded spatial profile" in Figure \ref{fig:psfs} x-axis as it effectively only shows half the slit due to their combination), such that the shown BA$_{14}$ profile is the aligned-and-stacked average of all 12 200-second exposures on that night. The spatial profiles of the star are from single combined A-B pairs due to the much higher SNR on the brighter calibration star. The profiles were extracted at wavelengths corresponding to the J-filter, or between 1.17 and 1.33 $\mu{m}$. The Full-Width Half-Max (FWHM) of a best-fit Gaussian profile to the stars retrieved FWHM$=0.28\arcsec \pm 0.01\arcsec$ before and FWHM$=0.28\arcsec \pm 0.01\arcsec$ after observations of the BA$_{14}$, which was measured to have FWHM$=0.39\arcsec \pm 0.01\arcsec$. Performing the same analyses at H or K band retrieved the same FWHMs for the stars ($0.26-0.28\arcsec$) and comet ($0.38-0.39\arcsec$) as for J to within $\sim1.0\sigma$.

The extended PSF of BA$_{14}$ in our IRTF observations could be due to an actual detection of its coma in the near-infrared (despite the generally much lower visibility of typically sized cometary dust at those wavelengths) or could be due to slight guiding imperfections that might not be noticed under ordinary seeing conditions. An object being tracked at non-sidereal rates ($1.4\arcsec/min$ on the 7th) with long exposure times will inevitably lead to some drift within the slit over the course of the exposures; both of these issues do not apply to the sidereally tracked short exposures on the nearby standard stars. If the detection is indicative of the coma's size, given the geocentric distance of $\Delta=0.125 AU$, we estimate the dust coma BA$_{14}$ to have a FWHM of $\sim35-36$ kilometers, and a maximum detectable extent of $\sim240$ kilometers at J-band on March 7th UTC. If the coma was the same size on March 10, we estimate it's FWHM to have been $\sim0.49\arcsec$, though we note that this is considerably smaller than the seeing limit on that date.

To assess the appearance of the coma at optical wavelengths around the same time, we queried the Mission Accessible Near-Earth Object Survey (MANOS)'s database and found visible-wavelength imaging observations from the SOAR telescope and Lowell Discovery Telescope (LDT) on 2016 February 21 and 22, respectively, when the heliocentric and geocentric distances of BA$_{14}$ were $R_H = 1.055-1.059 AU$ and $\Delta = 0.229-0.237 AU$. The SOAR observations utilized the Goodman High-Throughput Spectrograph \citep{2004SPIE.5492..331C} in imaging mode with the Sloan r filter, while the LDT observations utilized the Large Monolithic Imager (LMI, \citealt{2013AAS...22134502M}) with a "VR" filter. \edits{The SOAR observations consisted of one 10 second exposure and five 60 second exposures, while the LDT observations were six 60 second exposures.} We applied standard corrections to each set of images (debiasing, flattening) and then calibrated them astrometrically and photometrically using the \textit{Photometry Pipeline} \citep{2017A&C....18...47M}, with the photmetric calibration being derived exclusively from Sun-like stars within the fields of view. (The "VR" filter was calibrated as if it was a Sloan r filter due to their similar central wavelengths.)

In the LDT observations, the FWHM of BA$_{14}$ was $0.95\pm0.02\arcsec$ along the direction perpendicular to its apparent motion and $1.07\pm0.02$ along the direction of its motion compared to a uniform $0.91\pm0.01$ for nearby field stars. For the SOAR observations, the cross track profile has $0.51\pm0.01\arcsec$, the along-track profile has $0.59\pm0.02$, and nearby stars have $0.49\pm0.01\arcsec$. The images were all taken with exposure times short enough to prevent the along-track extension being due to trailing. \edits{These quoted numbers are measured from the composite images composed of the individual frames stacked at the object's non-sidereal rates, but the same results were found through inspection of individual frames as well.} Through these analyses and visual inspection of all frames individually and stacked, it seems rather firm that we detect BA$_{14}$'s small coma in these images. The magnitude of BA$_{14}$ was $m_{r} = 18.71 \pm 0.02$ on Feb. 21 and $m_{VR-r} = 18.64 \pm 0.02$. 

The visible coma of BA$_{14}$ was two-to-three times the size of the NIR spatial profile at twice the geocentric distance and when the comet was slightly further from the Sun and thus presumably slightly less active. If any of the slightly-larger-than-stellar spatial profile in our IRTF observations is from reflected or thermally emitted light from the coma of BA$_{14}$, it is sampling a small areal fraction of an already small coma at wavelengths where dust should be less prominent. In the Section \ref{sec:results}, we discuss the very limited possible coma contamination of our NIR spectra.

\begin{deluxetable*}{cccccCrlcc}[b!]
\tablecaption{Summary of Infrared Observational Data\label{tab:observations}}
\tablecolumns{9}
\tablenum{1}
\tablewidth{0pt}
\tablehead{
\colhead{UT Date\tablenotemark{a}} &
\colhead{UTC Time\tablenotemark{a}} &
\colhead{$R_H$} &
\colhead{$\Delta$} &
\colhead{$\alpha$} &
\colhead{Seeing} & 
\colhead{RH\tablenotemark{b}} & 
\colhead{Total Time}\\      
\colhead{(YYYY-mm-dd)} & 
\colhead{(d)} &
\colhead{(au)} &
\colhead{(au)} &
\colhead{$(^{\circ})$} &
\colhead{(arcsec)} & 
\colhead{\%} & 
\colhead{seconds}
}
\startdata
2016-03-07 & 05:26 & 1.015 & 0.125 & 77.0 & $\sim 0.51\arcsec$ & $\sim$ 22 & 2400.0  \\
2016-03-10 & 05:12 & 1.011 & 0.102 & 76.1 & $$\sim1.5\arcsec$ & $\sim$ 6  & 2400.0  \\
\enddata
\tablenotetext{a}{ UTC at start of sequence.}
\tablenotemark{b}{ Relative Humidity as measured at the IRTF at the start of sequence.}
\end{deluxetable*}

\section{BA$_{14}$ in the NIR} \label{sec:results}
\subsection{Retrieved Reflectance Spectra}
\begin{figure}[ht!]
\plotone{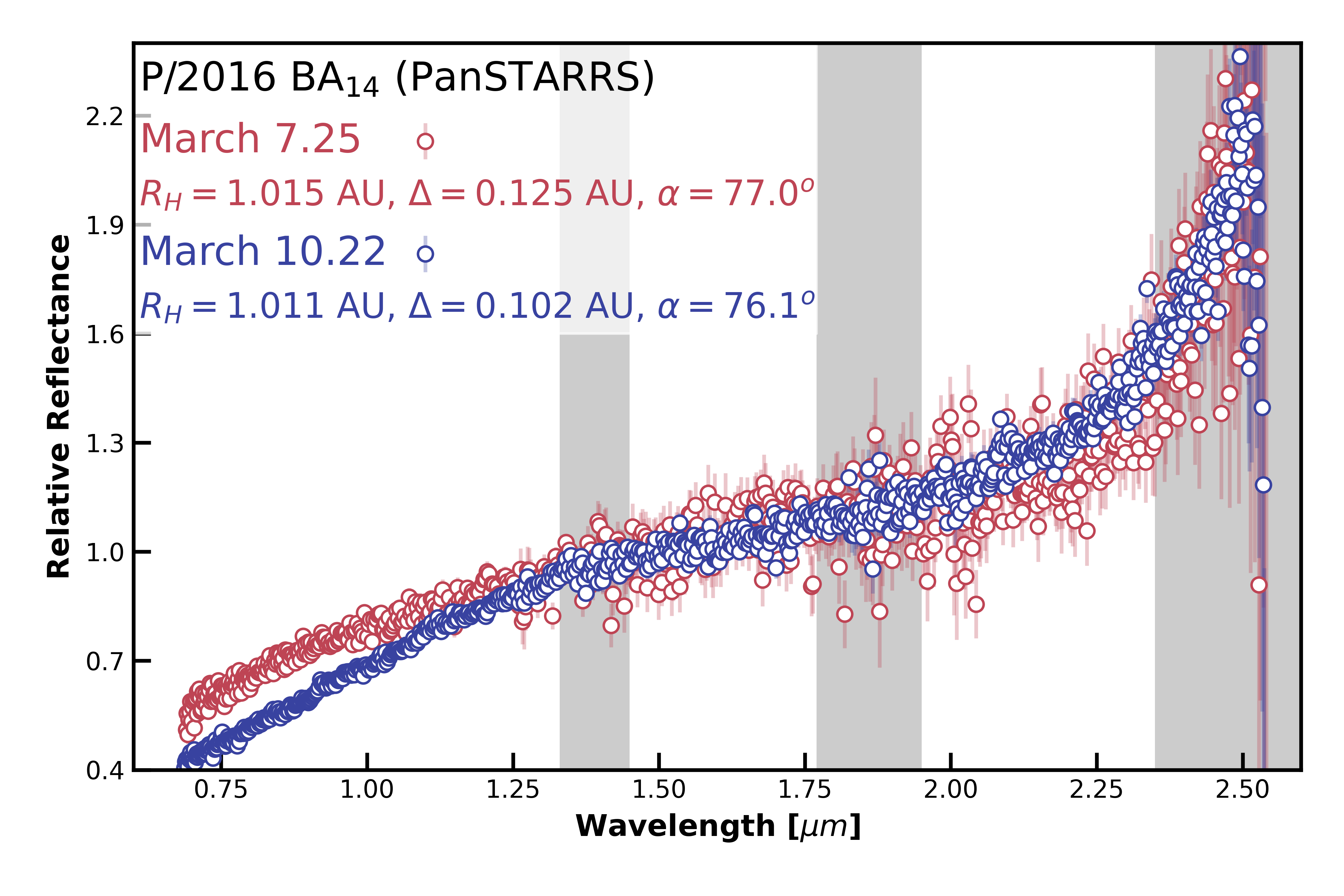}
\caption{The retrieved reflectance spectra of P/2016 BA$_{14}$ (PanSTARRS) as obtained on 2016 March 7 and 10 from the NASA IRTF, normalized such that $R(\lambda=1.5\mu{m})=1.0$. The spectra on both nights are similarly red-sloped with a thermal excess visible at longer wavelengths ($>2.25\mu{m}$, and the spectrum obtained on March 10th is slightly redder at all wavelengths with a slight slope-break near $1.4-1.5\mu{m}$ and a more linear spectrum at shorter wavelengths. See text for more details.}
\label{fig:spectra}
\end{figure}

The retrieved reflectance spectra of P/2016 BA$_{14}$ (PanSTARRS) are shown in Figure \ref{fig:spectra}. The spectra obtained are similar on both dates, with an overall red-slope becoming more neutral with wavelength and an upturn in reflectivity at wavelengths longer than $\sim2.25\mu{m}$ that is likely excess thermal emission "on top" of the reflected light from the solid coma and nucelus. However, the spectrum obtained on March 10 is noticably redder at shorter wavelengths becoming similar to the March 7 observations by $\sim$H-band, the possible slope-break near $1.4-1.5\mu{m}$ appears somewhat sharper, and the region $\lesssim1.35$ appears slightly more linear. Both stars were slope-calibrated to the same well-studied solar analog (SAO 93936) and inspection of the ratios of the spectra of each star's local G-type telluric would not have introduced the change either.We also note that any wavelength-dependent slit losses would be smooth functions of wavelength, unlike the difference in the retrieved spectra, so we think this an unlikely cause of the discrepancy as well -- especially considering that our observations were all at the local parallactic angle to mitigate just this effect. Considering that conditions on both nights were rather good (low and stable humidity), the phase angle did not change appreciably, and the target was observed at similar airmass on both nights, it seems like the difference is real, or at least mostly so.

To quantify the slopes of the two spectra, we use the $S'$ framework of \citet{1990Icar...86...69L} and report slopes for the J, H, and K spectral ranges in units of $\% / 0.1\mu m$ for both nights. For the March 7 data, we find $S_{J}'=4.6\pm0.8$, $S_{H}'=3.4\pm0.7$, and $S_{K*}'=4.3\pm1.5$. For March 10, we find $S_{J}'=7.7\pm0.3$, $S_{H}'=3.0\pm0.3$, and $S_{K*}'=5.9\pm0.6$. The K-slopes are contaminated from the thermal emission, and we report slopes measured over the range ($1.95-2.20\mu{m}$) to attempt to mitigate this. At wavelengths away from the thermal excess, the slopes of the two spectra agree at the $1-\sigma$ level at H-band, but March 10 is significantly redder at J-band as expected.

Furthermore, it is worth stating explicitly that our spectra are almost certainly quite phase-reddened \citep{2012Icar..220...36S}, meaning that the retrieved reflectance spectra appear redder than they would if observed at a lower phase angle. However, estimating the magnitude of this effect is challenging as we don't know the phase function of the nucleus or the dust coma, the relative contribution of either to the reflectance spectrum, or how that would vary with phase angle itself.

\subsection{Did we observe the nucleus?}
Perhaps the most obvious factor to consider, given the claimed direct observation of the nucleus by \citet{2021Icar..36314425O}, would be the $\sim20\%$ drop in geocentric distance between the two dates from $\Delta=0.125 AU$ on March 7 to $\Delta=0.102 AU$ on March 10th. One would expect, \textit{a priori}, that if the coma was more spread out on the sky then the brightness of the nucleus might dominate more in the later, closer observations. Making a quantitative prediction of \textit{how} much fainter a dust coma might be more straightforward at larger geocentric distances, but for the present case is likely to remain fairly speculative. Given the rather long rotation period ($P\sim40$ hours, \citealt{2016DPS....4821905N}) and the small scales probed, it seems reasonable to assume that the innermost coma of BA$_{14}$ is highly anisotropic unless the nucleus's surface is uniformly active, which seems rather unlikely given its extremely low active fraction \citep{2017AJ....154..136L}. The visible light curve of the object presented in \citep{2017AJ....154..136L} shows that the object's overall brightness trend is dominated by the nuclear signal, though this does not necessarily preclude some contamination.

The slight change in spectral behavior between the two observations also provides insight into the issue. The spectrum obtained on 2016 March 10, when coma contamination should have (in principle) be lower and the nucleus more prominent, shows a slope-break near $1.4-1.5\mu{m}$. A population of small grains comparable to the wavelengths of note (as expected for cometary comae) reflects light according to Mie theory, which should produce reflectance spectra that do not have discontinuities or sudden slope changes with wavelength but instead more gradual continuous ones. By contrast, light reflected by solid surfaces or by particles much larger than the wavelength of light in question, which are out of the Mie regime and better modeled with Hapke photometric theory \citep{1993tres.book.....H}, can produce such sudden slope breaks. In fact, \citet{2019AJ....158..204S} found a slope-break at a nearly-identical wavelength on the Jupiter Trojan and \textit{Lucy} \citep{2016LPI....47.2061L} mission target (21900) Orus. A review of the limited number of nuclear spectra of confirmed low-activity or dormant comets in \citet{2008Icar..194..436D} and \citet{2021PSJ.....2...31K} show that \textit{some} but not \textit{all} comets in those studies have slope breaks in the same wavelength region. Furthermore, the change in slope is very apparent on comet 67P/Churyumov-Gerasimenko \citep{2019SSRv..215...19F}. As to whether or not the slope-break derives from large grains or the surface of the nucleus itself, a nucleus with as low an active fraction as BA$_{14}$ ($\sim0.01\%$, \citealt{2017AJ....154..136L}) would make ejection of large grains in large number physically challenging. Furthermore, if the shorter wavelength section of March 7 dataset is more curved and slightly less red than the March 10 spectrum but the two spectra converge at longer wavelengths, one possible explanation is that the shorter wavelengths of the March 7 reflectance spectrum contain some light from grains similar in size to the wavelengths being studied here -- e.g., $\mu{m}$-sized, within the Mie regime and outside of the Hapke regime. We thus conclude that the most likely scenario to explain our observations is that the nucleus is a majority of the reflected light contained in the spectra collected on March 7 and March 10, and that the March 10 spectrum is likely even more dominated by the nucleus itself.

\section{Thermal Excess Modeling} \label{sec:thermal}
\subsection{Modeling Description and Initial Results}
Understanding, modeling, and correcting for the exponential-like upturn at wavelengths longer than $\sim2.25\mu{m}$ in spectra of low-albedo NEOs is critical to both understanding their reflectivities at those wavelengths as well as understand the thermal properties of their surface and near-surface. The go-to method for these kinds of studies is the Near-Earth Asteroid Thermal Model (NEATM, \citealt{1998Icar..131..291H}), a modification of the earlier Standard Thermal Model \citep{1986Icar...68..239L} to better suit observations of small NEOs. The specifics of applying NEATM to thermally-contaminated reflectance spectra at these and similar wavelengths is outlined in \citet{2005Icar..175..175R} and \citet{2009M&PS...44.1917R}. NEATM (and STM, for that matter) assume that all thermal emission from the object in question is blackbody emission from the dayside of the object. The surface temperature at a particular point is calculated as:
\begin{equation}
    T(\theta, \phi) = [ \frac{(1-A)\times S_\odot}{\eta\epsilon\sigma\times R_{H}^2} \times cos(\theta) \times cos(\phi) ]^{1/4}
\end{equation}
where $A$ is the Bond Albedo of the object (which in turn is a function of $p_V$), $S_\odot$ is the Solar Constant (1370 $W m^{-2}$ at $R_H=1$ AU), $\eta$ is the infrared beaming parameter, $\epsilon$ is the (bulk) emissivity of the object (assumed to be the canonical $0.9$), $\sigma$ is the Stefan-Boltzmann Constant, and $R_H$ is the heliocentric distance of the object at the time of observation in AU. $\theta$ and $\phi$ are the angular coordinates on the body itself moving away from the subsolar point, such that the temperature of the object is modeled as $T=0$ at the terminator and on the whole of the nightside. The infrared beaming parameter $\eta$ is essentially the 'fudge factor' that allows the model to overcome its rather large assumptions. A body with 'normal' (e.g. regolith covered and moderate-to-slow rotation speed around its prinicpal axis) thermal parameters which observed at rather small phase angle should have a value of $\eta\sim0.9-1.1$ -- only a small deviation from a traditional greybody, but at larger phase angles or with less common thermal states (less regolith, non-principal axis rotation, etc.), the value of $\eta$ should get increasingly large, up to a (theoretical) maximum near $\pi$. While a more detailed description of the model and its serious limitations in this capacity is included in \citet{2018AJ....156..287K}, we note that as the model is really constraining $T_{SS}$, the fit parameters $p_V$ and $\eta$ are expected to be somewhat degenerate.

\begin{figure}[ht!]
\plotone{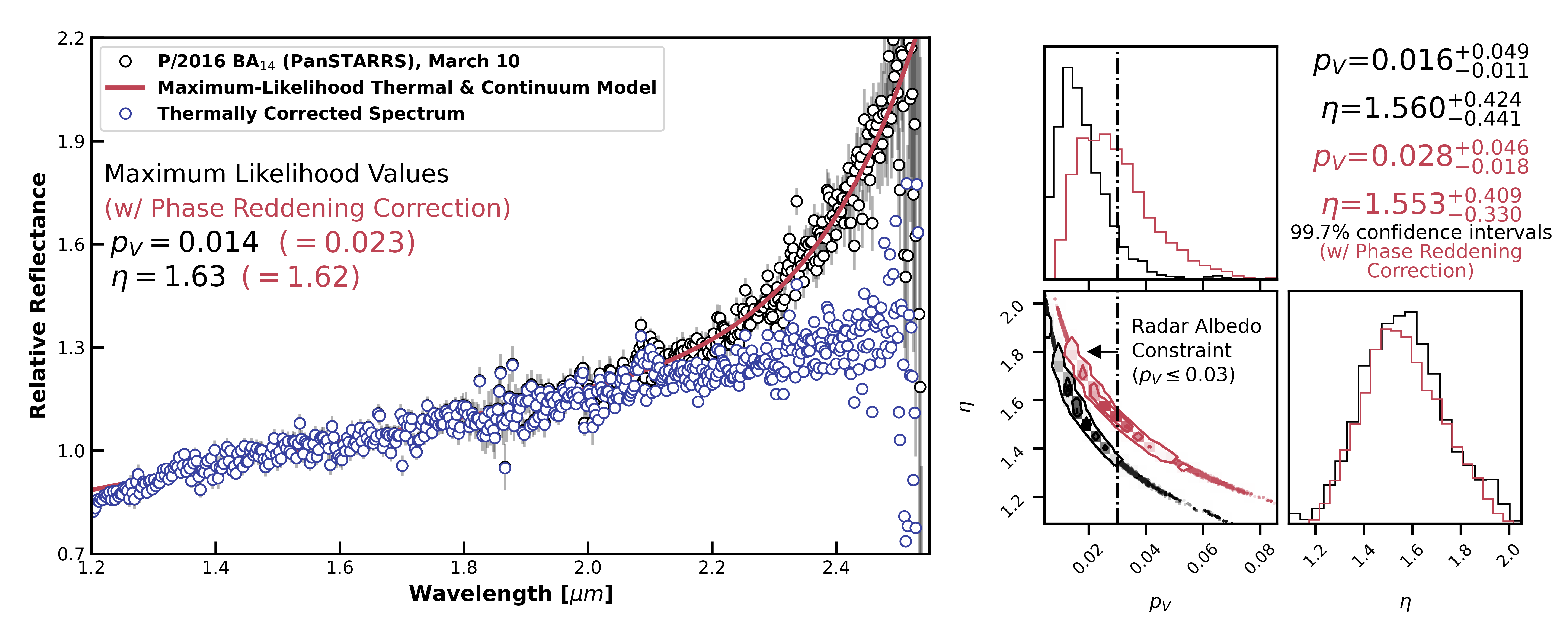}
\caption{Left: The Maximum Likelihood Thermal Model, with $p_V=0.014$ and $\eta=1.627$, or $p_V=0.023$ and $\eta=1.622$ after a correction for phase reddening), is shown as a dark red line, while the uncorrected data is shown as black unfilled circles. The thermal-model-corrected all-reflectance spectrum is shown as blue unfilled circles. Right: A corner plot shows the distribution of fit parameter values for $p_V$ and $\eta$ for both models. The albedo constraint from radar observations \citep{2016DPS....4821905N} is plotted as a vertical line. As expected, the albedo and beaming parameters show some degeneracy, as the model constrains the peak subsolar temperature which is a function of both parameters. (See text for more details.)}
\label{fig:models}
\end{figure}

In order to account for this degeneracy and better understand the errors on our fit parameters, we update the model of \citet{2018AJ....156..287K} by changing it from a least-squares fitting technique into a Markov-Chain Monte Carlo technique using the package \textit{emcee} \citep{2013PASP..125..306F}. The best-fit from the least-squares technique is used as an initial guess for the distribution of walkers in \textit{emcee}. We utilized 32 walkers and 2000 iterations as an initial attempt due to the low amount of parameters and length it took for each iteration. The results are presented in Figure \ref{fig:models}.

The fit parameters are constrained to be in the range of $p_V = 0.016_{-0.011}^{+0.049}$ and $\eta = 1.560_{-0.441}^{+0.424}$ at the 99.7$\%$ confidence level, with the Maximum-Likelihood values being $p_V=0.0135$ and $\eta=1.627$. The shapes of the distributions of the parameters are as expected, with the bottom left panel of the corner plot (the right side of Figure \ref{fig:spectra}) showing a curve. Essentially, this is the curve of constant peak subsolar temperature. While the marginalized distribution of $\eta$ values is symmetric about its approximate maximum likelihood value, the distrubtion of $p_V$ values is highly skewed, with the maximum likelihood value being on the absolute lowest end of the confidence interval.

These values for the infared beaming parameter $\eta$ are quite reasonable for a typical NEO, in fact almost exactly what would be estimated from the \citet{2003Icar..166..116D} empirical relationship between phase angle and beaming parameter. The range of allowed visible albedos are all really quite low -- if taken at face value, it would make BA$_{14}$ among the darkest objects in the Solar System. The radar data do mandate that the object's albedo has to be below 3$\%$, so our range of values do satisfy the one pre-existing constraint. The distribution for NEOs as found by the all-cryogenic phase of the WISE mission \citep{2016AJ....152...79W} finds that approximately $5\%$ of objects have geometric albedos of $0.02$ or less, so our nominal range of values place the albedo of $BA_{14}$ in the 'quite rare but not implausible' range. The radar albedo upper limit and an absolute magnitude of $H_V = 19.2$ result in a diameter estimate of $D\sim1.1$ km, and our maximum-likelihood albedo of $p_V=0.0135$ results in a diameter of $D\sim1.6$ km. In the next two sub-sections, we discuss complications to this albedo estimate that would likely result in it being a slight under-estimate, and thus our estimated diameters being a slight over-estimate.

\subsection{Phase-Reddening Complications and Correction}
In order to estimate the ratio $f_{IR}=p_{IR}/p_{V}$, we extrapolated downwards to $0.55\mu{m}$ by way of a linear fit to all datapoints short of $0.9\mu{m}$. This is the data-driven way to estimate $f_{IR}$, but is susceptible to over-or-under-estimating the actual ratio should there be slope-changes in the visible. A linear extrapolation of the March 10 data results in $f_{IR}=3.24$. In many cases, one can check this estimate by comparing the retrieved value against typical values for the taxonomic class of the object, most commonly the Bus-Demeo System \citep{2009Icar..202..160D}, but our spectrum is even redder than the very-red A and D types (before phase reddening is accounted for, at the very least.) For context, the low-activity Halley-Type Comet P/2006 HR$_{30}$ (Siding Spring) had $f_{IR}\sim1.92$ \citep{2021PSJ.....2...31K} and the more active Jupiter Family Comet 67P/Churyumov-Gerasimenko had $\sim2.0$ depending on which area of the surface was studied \citep{2016Icar..272...32Q}. The retrieved spectrum of $BA_{14}$ itself is almost certainly significantly phase-reddened, which we first highlighted in Section \ref{sec:results}. The reflectance spectra of Solar System objects appear redder at larger phase angles than they would at zero-phase (e.g. opposition). While the magnitude of this change \textit{per degree} can be quite small (often smaller than the other measurement uncertainties at hand), at the large phase angles sometimes occupied by NEOs, it can be critical to account for it to estimate what the 'real' spectrum looks like. 

In the case of phase-reddened and \textit{hot} NEOs, the thermal emission at longer wavelengths is being added onto a reddened underlying continuum, making guesses at the underlying reflectivity at longer wavelengths challenging. Phase reddening is not constant with wavelength, so there is some question of how much of the thermal upturn is actual thermal emission and how much is the phase-induced redness of the object underneath the thermal photons. There has been some work done to correct NEATM's imperfections at larger phase angles (e.g. \citealt{2018AJ....155...74M} calculates corrections to NEATM for high phase angles, but their work assumes all light being collected is flux-calibrated observations at longer wavelengths than studied here), but we are not aware of a pre-existing prescription in the literature for dealing with this particular issue.

Given the lack of additional information, we attempt to "undo" the phase reddening's impact on our thermal modeling by re-running our MCMC modeling routine with a lower $f_{IR} = 1.92$, implicitly assuming that BA$_{14}$'s surface is similarly red (measured in bulk from the visible to K-band) at zero phase angle to P/2006 HR$_{30}$ (Siding Spring)'s. These results are plotted in red on Figure \ref{fig:models}. The distribution of $\eta$ is essentially unchanged ($\eta = 1.553_{-0.330}^{+0.409}$), but the distribution of visible geometric albedo $p_V$ is shifted higher ($p_V = 0.028_{-0.018}^{+0.046}$) by approximately the ratio of $f_{IR}$ values ($3.24/1.92 \sim 1.7 \sim 0.28/0.016$). This is the expected behavior for small enough changes in $f_{IR}$: the actual thermal model can fit the data just as well (in this case, it overlaps nearly exactly), it is the distribution of plausible visible albedos that is modified.

So, which distribution of albedos is more "likely" for this object? There is no obvious answer without a complementary visible spectrum or broadband color measurement or phase reddening study of this object. It seems \textit{likely} that BA$_{14}$'s visible colors are similar to that of HR$_{30}$ or 67P, and it seems inevitable that BA$_{14}$'s spectrum is at least somewhat phase reddened, so some correction is definitely in order, but the "correct" magnitude of such a correction is unknowable at present. In any case, the quality of the data doesn't allow an obvious discrimination between such small changes in albedo anyways. The albedo of BA$_{14}$ is almost certainly between $0.01-0.03$, but precision beyond that would require higher signal to noise thermal observations at longer wavelengths -- or, of course, a visit by \textit{Comet Interceptor.} More constraints from other techniques would not allow us to constrain the albedo better, but it would allow us to model the thermal emission more accurately and better understand the underlying reflectance spectrum with more confidence.

In summary, our thermal models of BA$_{14}$'s thermal emission naturally agree with the $p_V \leq 0.03$ of \citet{2016DPS....4821905N} regardless of a phase-reddening correction, the distributions of beaming parameters $\eta$ seem entirely typical for a NEO observed in this geometry, and the underlying spectrum appears to be featureless and linear from $1.4\mu{m}$ onwards like many other comets, including 67P.

\subsection{Possible Thermal Contamination By Dust}
Despite BA$_{14}$'s thin-and-small coma, the dust grains that are there would still be quite warm at $R_H = 1.01$ AU, suggesting that some of the thermal emission that we model as coming from the surface of the object \textit{could} be coming from the coma instead. If the coma of BA$_{14}$ is composed at least partially of rather small grains only a few $\mu{m}$ in size -- which seems natural for an object with such weak activity -- the grains in the coma might appears "hotter" than the nucleus as they cannot effectively emit at wavelengths larger than their physical size (see, e.g.,  \citealt{2004ApJ...612..576S}). This process, often called "superheating", would further bias the retrieved thermal fits towards higher temperatures and thus lower albedos and beaming parameters. However, BA$_{14}$'s activity is quite weak, the quality of the fits are rather good, and the retrieved range of values for the infrared beaming parameter $\eta$ are quite reasonable for a small NEO. If there were significant 'extra' thermal emission, then the shape of the blackbody \edits{would} change and become steeper -- the fact that we were able to fit it reasonably implies that the subsolar temperature our reflected-light-and-thermal models used was reasonable to correct the data \textit{assuming that the thermal emission and reflected light were coming from the same source}. In principle, we might be unable to differentiate our all-nucleus model from a stranger situation, such as one with a higher albedo nucleus and a darker albedo dust that balance such that their composite thermal emission looks like the curve we've modeled here. However, we would expect the dust coming off the nucleus's surface to look similar to the surface itself, and thus we argue that, again, this is likely not a huge factor for this dataset.\footnote{We remind the reader here that there are shortcomings of NEATM in general, summarized in \citet{2018AJ....156..287K} and elsewhere.} We also note that if the contamination by dust was a large factor, it seems likely that our model would not be able to satisfy the albedo constraint imposed by the radar albedo. As with phase reddening, this might result in a slight \textit{under-estimate} of the true albedo of BA$_{14}$ if it were significant.

\section{Discussion} \label{sec:discussion}
The first reason that we were interested in re-examining our spectra of P/2016 BA$_{14}$ (PanSTARRS) was to compare with the recent work of \citet{2021Icar..36314425O}. If BA$_{14}$ looked strange or anomalous at mid-infrared wavelengths, what would it look like at other wavelengths? Do the reported mid-IR phyllosilicate and organic absorption features reported for BA$_{14}$'s nucleus manifest with any signatures in the near-IR? If BA$_{14}$ looks like a 'typical' comet using these other approaches, how does one interpret the general applicability of the \citet{2021Icar..36314425O} result? The second reason was the selection of BA$_{14}$ as a back-up target for \textit{Comet Interceptor} \citep{2020RNAAS...4...21S} should a suitable Long Period Comet (or Interstellar Object!) not be found in time. In this section, we first compare and contextualize our results with those of \citet{2021Icar..36314425O} and then discuss its viability as a spacecraft mission target afterwards.

\subsection{Comparison to the Mid-Infrared}
The reflectance spectra we obtained of BA$_{14}$ in this work on both nights appear very much 'typical' for comets observed in the near-infrared when phase-reddening is considered. We interpreted our observations in Section \ref{sec:results} as both being primarily dominated by light from the nucleus itself, with the March 10 spectrum being even more so. The different spectral behaviors on the two nights are consistent with a thin dust coma around the object object composed of $\mu{m}$-sized grains, which is what would be expected for such a weakly active object \citep{2017AJ....154..136L}. Our thermal model estimates the visible-wavelength albedo of BA$_{14}$ to be quite low at $p_V = 0.016_{-0.011}^{+0.049}$ at the $99.7\%$ formal confidence interval, though the large phase angle and continuing low-level dust contamination both likely result in this being a slight under-estimate of the albedo of the nucleus. A simple phase-reddening correction results in a slightly higher modeled albedo distribution of $p_V = 0.028_{-0.018}^{+0.046}$ ($99.7\%$ confidence interval). Radar data from that apparition \citep{2016DPS....4821905N} suggest that the albedo has to be below $0.03$, and there are cometary nuclei whose albedos are within this range, like 19P/Borrelly \citep{2002Sci...296.1087S} or 103P/Hartley 2 \citep{2009PASP..121..968L}. If our albedo estimates are accurate, then we estimate a diameter between $1.1-1.6$ km, again consistent with the current understanding of the radar data, and very much a typical size for a Jupiter Family Comet nucleus \citep{2017AJ....154...53B}. In the unlikely case that the object's albedo really is on the low end of our modeled albedo distributions, the diameter of $BA_{14}$ could be up to $\sim2$ km across. The infrared beaming parameter modeled here ($\eta = 1.560_{-0.441}^{+0.424}$) is a typical range of values for an NEO observed at similarly high phase angles as is the case in this study \citep{2003Icar..166..116D}. In summary, BA$_{14}$ very much appears to have the properties expected for a normal comet at these wavelengths, though it might be of slightly lower albedo than many cometary nuclei (pending the previous discussions of phase reddening and dust, of course). The only truly rare traits it seems to have are its low activity levels, its slow rotation state, and its anomalous appearance at mid-infrared wavelengths.

While the mid-infrared spectrum of BA$_{14}$ suggests a surface incorporating more processed substances like phyllosilicates, the near-infrared nuclear spectrum does not appear different than other comets expected to have a more typical composition. \citet{2021Icar..36314425O} attribute several features that they cannot identify to being from a matrix of organics, which is thought to be a dominant surface material in comets like 67P/C-G and others \citep{2019SSRv..215...19F}. One possibility is simply that the matrix of complex organics is the dominant factor in determining the visible and near-infrared spectral properties of the nucleus (considering that cometary nuclei are dark and red at those wavelengths, this seems plausible) but the balance changes as one observes at longer and longer wavelengths. Could BA$_{14}$ have a different surface composition compared to other objects, even if NIR spectroscopy cannot discern it? \citet{2021Icar..36314425O} argue based on the apparent similarity of their spectrum to heated micrometeorites that the difference is evolutionary and likely driven by previous high-temperature conditions brought on by a past orbit with a lower perihelion. However, \citet{2021PSJ.....2...31K} found some evidence that cometary objects should have spectral changes in the visible/near-infrared range if they had been heated significantly, so the fact that BA$_{14}$ looks like a 'typical' comet in the NIR does not necessarily support a heating origin, but more data are needed on these kinds of objects to say with any more certainty. A future detailed study of BA$_{14}$'s and 252P's dynamical evolution is certainly warranted, both to understand their mutual dynamical relationship should it be robust and to better understand BA$_{14}$'s past thermal state to assess the viability of the hypothesis that its modern surface is thermally altered.

Comets with decreasing activity levels (e.g. lowering gas production rates) can only lift smaller and smaller grains with time, in principle leading to a preferential loss of small grains from the surface. If this is a reasonable description for the recent evolution of BA$_{14}$'s surface, then it might be the loss of these small and porous grains which might normally produce the 10$\mu{m}$ excess is what allowed the observers to see 'through' to the underlying material which is usually buried in the mid-IR data. The study of \citet{2021PSJ.....2...31K} found that the nearly-dormant Halley-Type Comet P/2006 HR$_{30}$ (Siding Spring) had retained a classic comet/D-type slope despite whatever processes are ongoing as it became less active. That object and BA$_{14}$ both have perihelia similar to or greater than 1 AU, suggesting that they're also in similar thermal regimes, at least in their current and recent orbits.

One key problem going forward is the lack of clear mid-IR observations of many cometary nuclei and associated objects to compare and contextualize BA$_{14}$ and related objects in. Even low-level comae might be able to hide the kinds of surface features that \citet{2021Icar..36314425O} observed, assuming that they (and we!) were successful in actually observing the nucleus. One path forward is to obtain more $10\mu{m}$ observations of dormant comet candidates and low-activity comets in general to compare to and better estimate which attributes of objects like BA$_{14}$ are evolutionary and which parts are simply challenging to observe because of coma interference (a preliminary study reported in \citet{2018DPS....5031201M} suggests this might be quite fruitful.). \edits{The recent (successful) launch of JWST will facilitate a generational breakthrough in our ability to measure near- and mid-infrared reflectivities and emissivities of the surfaces of Solar System Small Bodies. This is doubly true for cometary surfaces, for which observations at larger heliocentric distances are usually required to mitigate the confusing effects of dust and gas emission and discern the properties of the surface more directly. The NIRSPEC (up to $5\mu{m}$) and MIRI ($5.0-28.5\mu{m}$) instruments are likely to be the instruments of choice, though the infrared imager NIRCAM might be better suited to fainter targets. At present, we are not aware of any currently-approved JWST programs to use these mid-infrared wavelengths to better understand the composition of low-activity/dormant comet nuclei, but it seems like a highly fruitful avenue for future research. Approved studies of the Trojan asteroids and the Centaurs, both populations thought to share significant similarities with the objects discussed here, should be highly useful for planning future low-activity comet studies in addition to their obvious scientific benefits.} Another hopeful note is the L'TES instrument on the Lucy mission will be able to measure the surface properties of more trojan asteroids in more detail, which have previously been shown to have mid-infrared spectra like that of 'typical' (non-BA$_{14}$-like) comets. Lastly, we also note that many datasets obtained of BA$_{14}$ during its 2016 close approach have yet to be fully analyzed and could provide key insights to help better understand how this object fits into the population of the comets in general.

\subsection{Comet Interceptor}
Among the back-up targets for \textit{Comet Interceptor} listed in \citet{2020RNAAS...4...21S}, BA$_{14}$ is one of two low-activity targets along with 289P/Blanpain. Low activity targets naturally allow for safer flybys with the possibility of closer approaches to the nucleus of the chosen comet, but also would result in less material being encountered and identified by the remote sensing instruments as well as (presumably) a less complex plasma environment. BA$_{14}$ does put off some gas \citep{2017AJ....154..136L}, so there is no reason to think this object would be strictly a bare nucleus, but the focus of a spacecraft visit would simply have to be more focused on studies of the surface of the comet. The slow rotation of the object as determined by radar when combined with the lower activity state could allow for higher resolution imaging and mapping of the nucleus than many of the other targets. There are a variety of interesting questions that could be addressed through detailed imaging and spectroscopy of a nearly dormant comet, such as "how does the bulk topographic roughness evolve as comets age?" or "can the activity history of a nearly/dormant comet be determined through detailed mapping of its surface?" Discriminating between surface properties (reflectance spectra, topography, thermal characteristics) between the small fraction of the surface that is active at present and the inactive fraction can only be accomplished through spatially resolved studies. If an understanding can be developed of \textit{why} certain areas are still active but others aren't (and how that might affect other observable features), then a more general purpose understanding of how comets age can be developed in a much more nuanced way than just assuming that their gas production rates trend towards zero with time. The dormant comets constitute at least a few percent of the NEO population, and understanding their material and structural properties would thus benefit studies of similar objects from a planetary defense point of view in addition to developing a broader understanding of cometary life cycles. The flyby of BA$_{14}$, a 'traditional' comet that has gone dormant, would provide a fascinating counterpoint to the upcoming flyby of (3200) Phaethon, an extremely low activity object that is quite unlikely to be a 'traditional' comet as opposed to a former more active asteroid, by JAXA's \textit{DESTINY$^+$} \citep{2018LPI....49.2570A}.

The MIRMIS instrument \citep{2020LPI....51.2097B} onboard Spacecraft A of \textit{Comet Interceptor} could produce exceptionally interesting results at characterizing BA$_{14}$. MIRMIS is a near-and-mid-IR spectrograph that would facilitate a detailed mapping of the cometary surface and would be able to near-simulatenously observe at the wavelengths of the data presented here and the mid-IR observations of \citet{2021Icar..36314425O} in a spatially resolved way across much of the nucleus. This could more clearly reveal what compounds are most dominant in determining the typically featureless red spectra of cometary nuclei at visible and near-infrared wavelengths through detecting their absorption features directly throughout the mid-IR as in \citet{2021Icar..36314425O}. The sensitivity of the instrument to multiple absorption features of water ice would also allow a more direct and sensitive measurement of its modern near-surface volatile content than would be capable from the ground. A comparison between the thermal stability of the surface materials, their distribution across the surface, and the orbital history of the object could then also directly assess how each of those materials responds to heating and thermal degradation once the cooling effect of sublimation is minimized. This would be a more direct test of \citet{2021Icar..36314425O}'s suggestion that BA$_{14}$'s nucleus shows signs of past more intense heating than we were able to achieve with our observations.

\section{Summary} \label{sec:summary}
We observed the low-activity Jupiter Family Comet P/2016 BA$_{14}$ (PanSTARRS) on March 7 and March 10, 2016 using SpeX on the NASA IRTF some days before its closest approach to the Earth later that month as part of an ongoing program to obtain reflectance spectra of close-approaching and notable NEOs. We were prompted to re-reduce and re-inspect the dataset after the recent publication of \citet{2021Icar..36314425O}, which found BA$_{14}$ to have an atypical spectrum at mid-infrared wavelengths compared to other comets. In particular, they interpret their observations as being directly of the nucleus of the object and for it to be covered primarily by phyllosilicates, as opposed to anhydrous materials more commonly found in cometary dust comae. The selection of BA$_{14}$ as a backup target for the ESA's \textit{Comet Interceptor} \citep{2020RNAAS...4...21S} provides another pressing reason to reinvestigate this intriguing object now. In particular, we sought to infer whether or not our NIR observations also probed the properties of the nucleus and how it compared to other comets observed in the NIR in general. We found:
\begin{itemize}
    \item The reflectance spectrum of BA$_{14}$ on both nights is steeply red and shows a significant thermal upturn at longer wavelengths. The March 7 spectrum, observed at a geocentric distance of $\Delta=0.125$ AU, has a more slowly varying spectrum, while the March 10 spectrum ($\Delta=0.102$ AU) has a sharper change in slope near $\sim1.4-1.5\mu{m}$. We interpret both spectra as being nucleus dominated through studies of their spatial profiles compared to visible observations from the same apparition. The latter spectrum is likely even further dominated by the nucleus due to the lower surface brightness of the coma being spread out over a larger area of the sky and the challenges with producing such a spectrum from a population of grains as opposed to a solid surface. The slope break at $1.4-1.5\mu{m}$ is spectrally similar to 67P \citep{2016Icar..272...32Q}, but at a longer wavelength to the slope break on comets like P/2006 HR$_{30}$ (Siding Spring) \citep{2021PSJ.....2...31K}.
    \item We employ the Near-Earth Asteroid Thermal Model (NEATM) using a new MCMC-based implementation to our March 10 spectrum to constrain the visible wavelength albedo ($p_V$) and the near-infrared beaming parameter ($\eta$). At the $99.7\%$ confidence level and assuming typical values for emissivity, the slope parameter G, etc., we retrieve $p_V = 0.016_{-0.011}^{+0.049}$ and $\eta = 1.560_{-0.441}^{+0.424}$. This visible wavelength albedo is low, but within line for some comets visited by spacecraft and a $\sim$ few $\%$ of NEOs. The beaming parameter values are what would be expected for an object observed at these phase angles. Thermal contamination by dust and the large phase-reddening expected for this object make these albedo estimates slightly too low. As a result, we re-ran our thermal models accounting for phase reddening by assuming it had a typical cometary surface and retrieved $p_V = 0.028_{-0.018}^{+0.046}$ also at a $99.7\%$ confidence interval. Radar observations \citep{2016DPS....4821905N} constrain the albedo to be $p_V \leq 0.03$, which our models naturally reproduce.
    \item Assuming our nominal albedo range is accurate, we estimate the diameter of BA$_{14}$ to be $D\sim1.1-1.6$ km, which would be a typical size for a Jupiter Family Comet. The reflectance properties at near-infrared wavelengths of BA$_{14}$ (e.g., similar to many previously observed cometary nuclei) are seemingly at odds with a rare surface composition suggested by the mid-infrared observations of \citet{2021Icar..36314425O}. The two could be reconciled if VIS/NIR spectral response of BA$_{14}$ is minimally effected by the phyllosilicates that appear more prominent in the mid-infrared and more dominated by the low-albedo featureless matrix of organics that appears to be a common surface material on other comets.
    \item While it is tempting to infer that BA$_{14}$'s mid-IR derived surface composition is evolutionary, we note that the observational circumstances comparable to those for BA$_{14}$'s close approach are rare and high SNR observations of cometary nuclei even more so, so it cannot be ruled out that more 'typical' comets might look as BA$_{14}$ did in the near- and mid-IR should a similar circumstance present itself. If its properties are evolutionary, we argue for a slow loss of smaller grains with diminishing gas production as a more plausible mechanism to explain its surface composition as opposed to thermal alteration from a past hotter orbit, but more observations are needed.
\end{itemize}

We encourage observers who obtained data on BA$_{14}$ \edits{and its purported sibling 252P/LINEAR} during \edits{their} 2016 close approach but have yet to present it formally somewhere to re/analyze their observations to help understand \edits{these} compelling objects and where \edits{they} fit within the continuum of low activity and nearly dormant comets \edits{and their higher activity (possible) siblings.} We also support continued observation of low activity and dormant comet candidates at near- and mid-infrared wavelengths to add to the limited number of well-studied objects. Finally, a full dynamical study of the orbital evolution of P/2016 BA$_{14}$ (PanSTARRS) and 252P would be critical to understanding the mutual relationship between the two bodies and understanding the role that orbital evolution might have played in BA$_{14}$'s modern near-dormancy. All of these observations would also help the \textit{Comet Interceptor} team make an informed decision if their ideal comet is not found in time.

\begin{acknowledgments}
The MCMC thermal model described and implemented here is available upon reasonable request to the authors, and is being developed for public release in 2022.

The authors wish to recognize and acknowledge the very significant cultural role and reverence that the summit of Mauna Kea has always had within the indigenous Hawaiian community. We are most fortunate to have the opportunity to conduct observations from this mountain.

This work made use of observations obtained by the Mission Accessible Near Earth Object Survey (MANOS) which is funded through the NASA NEOO program grants NNX14AN82G and NNX17AH06G.
\end{acknowledgments}

%

\vspace{5mm}
\facilities{IRTF(SpeX)}


\software{NumPy \citep{harris_array_2020}, SciPy \citep{virtanen_scipy_2020}, AstroPy \citep{astropy_collaboration_astropy_2013,astropy_collaboration_astropy_2018}, spextool \citep{2004PASP..116..362C}, emcee \cite{2013PASP..125..306F}}




\bibliographystyle{aasjournal}



\end{document}